\documentclass[a4paper,aps,reprint,superscriptaddress,floatfix,longbibliography]{revtex4-1}

\usepackage{soul}   
\usepackage{graphicx}		
\usepackage{dcolumn}		
\usepackage{bm}				

\usepackage{amsmath}
\usepackage[usenames,dvipsnames]{xcolor}
\usepackage{microtype}
\usepackage{times}
\usepackage{changes}
\usepackage{hyperref}

\hypersetup{
  colorlinks,
  allcolors=NavyBlue,
  linktoc=all
}

\newcommand*\dagg{^{\dagger}}
\newcommand*\C{\mathcal C}
\newcommand*\heating{\alpha}
\newcommand*\omegacav{\omega\s{cav}}
\newcommand*\mat[1]{\begin{pmatrix}#1\end{pmatrix}}

\usepackage{ulem}
\newcommand*\matr[1]{\mathsf{#1}}
\newcommand{\unit}[1]{\ensuremath{\,\mathrm{#1}}}
\newcommand{\s}[1]{\ensuremath{{}_\mathrm{#1}}}	
\newcommand{\avg}[1]{\left\langle{#1}\right\rangle}

\begin{document}

\preprint{APS/123-QED}

\title{Floquet dynamics in quantum measurement of mechanical motion}
\author{Liu Qiu}\thanks{These two authors contributed equally}
\author{Itay Shomroni}\thanks{These two authors contributed equally}
\author{Marie A. Ioannou}
\thanks{Present address: GAP-Quantum Technologies, Universit\'e de Gen\`eve, Geneva 1211, Switzerland}
\author{Nicolas Piro}
\affiliation{Institute of Physics, \'Ecole Polytechnique F\'ed\'erale de Lausanne, Lausanne 1015, Switzerland}
\author{Daniel Malz}
\author{Andreas Nunnenkamp}
\affiliation{Cavendish Laboratory, University of Cambridge, Cambridge CB3 0HE, United Kingdom}
\author{Tobias J. Kippenberg}
\email{tobias.kippenberg@epfl.ch}
\affiliation{Institute of Physics, \'Ecole Polytechnique F\'ed\'erale de Lausanne, Lausanne 1015, Switzerland}

\date{\today}

\begin{abstract}
The radiation-pressure interaction between one or more laser fields and a mechanical oscillator gives rise to a wide range of phenomena: from sideband cooling and backaction-evading measurements to pondermotive and mechanical squeezing to entanglement and motional sideband asymmetry.
In many protocols, such as dissipative mechanical squeezing, multiple lasers are utilized, giving rise to periodically driven optomechanical systems.
Here we show that in this case, Floquet dynamics can arise
due to presence of Kerr-type nonlinearities, which are ubiqitious in optomechanical systems.
Specifically, employing multiple probe tones, we perform
sideband asymmetry measurements, a macroscopic quantum effect, on a silicon optomechanical crystal sideband-cooled to 40\% ground-state occupation.
We show that the Floquet dynamics, resulting from the presence of multiple pump tones, gives rise to an artificially modified motional sideband asymmetry by redistributing thermal and quantum fluctuations among the initially independently scattered thermomechanical sidebands. For pump tones exhibiting large frequency separation, the dynamics is suppressed and accurate quantum noise thermometry demonstrated.
We develop a theoretical model based on Floquet theory that accurately describes our observations.
The resulting dynamics can be understood as resulting from a synthetic gauge field among the Fourier modes, which is created by the phase lag of the Kerr-type response.
This novel phenomenon has wide-ranging implications for schemes utilizing several pumping tones, as commonly employed in backaction-evading measurements, dissipative optical squeezing, dissipative mechanical squeezing and quantum noise thermometry.
Our observation may equally well be used for optomechanical Floquet engineering, e.g.\ generation of topological phases of sound by periodic time-modulation.
\end{abstract}

\pacs{Valid PACS appear here}

\maketitle


\section{Introduction}
\label{sec:intro}
Cavity optomechanical techniques have enabled the preparation of mechanical oscillators close to the ground state via optomechanical sideband cooling~\cite{schliesser_resolved-sideband_2008,chan_laser_2011,verhagen_quantum-coherent_2012}, and therefore reach a regime where the interaction of mechanical oscillators and optical (or microwave) fields has to be treated quantum mechanically.
A wide range of quantum optomechanical phenomena can be probed by measuring the output spectrum in either homodyne or heterodyne detection, for example mechanical squeezing~\cite{kronwald_arbitrarily_2013, lecocq_quantum_2015,wollman_quantum_2015,pirkkalainen_squeezing_2015}, optomechanical squeezing~\cite{kronwald_dissipative_2014,Brooks2012,Safavi-Naeini2013,Purdy2014}, mechancial entanglement~\cite{woolley_two-mode_2014,ockeloen-korppi_stabilized_2018}, motional sideband asymmetry~\cite{safavi-naeini_observation_2012,purdy_optomechanical_2015,underwood_measurement_2015,weinstein_observation_2014, sudhir_appearance_2017}, as well as quantum correlations for variational measurements~\cite{Sudhir_Quantum_2017,Kampel_Improving_2017}.
Thus, precise knowledge and precision measurements of the sources and processes that contribute to the noise spectrum are imperative, such as thermorefractive noise~\cite{Braginsky2000,Anetsberger2011} and noise in drive tones~\cite{jayich_cryogenic_2012, safavi-naeini_laser_2013, kippenberg_phase_2013, Rabl2009}.

Here we demonstrate experimentally and describe theoretically how the Kerr-type nonlinearities, \textit{i.e.}~light-induced cavity frequency shifts~\cite{verhagen_quantum-coherent_2012,barclay_nonlinear_2005,suh_thermally_2012}, can lead to Floquet dynamics in the presence of multiple drive tones, coupling of originally independent thermomechanical sidebands (e.g. Stokes and anti-Stokes sidebands) and modification of the scattering rates.
This mechanism affects quantum measurements of mechanical motion, and we specifically demonstrate it leads to an artificially modified quantum sideband asymmetry.
When separating the driving tones in frequency far beyond the bandwidth of the Kerr-type
nonlinearity, the intrinsic motional sideband asymmetry is restored, and enables performing self-calibrated thermometry.
This mechanism is equally relevant to any other protocol that interrogates a cavity with multiple tones, such as backaction-evading measurements~\cite{suh_mechanically_2014,shomroni2018}, quantum squeezing~\cite{kronwald_arbitrarily_2013, lecocq_quantum_2015,wollman_quantum_2015,pirkkalainen_squeezing_2015}, entanglement of mechanical oscillators~\cite{woolley_two-mode_2014,ockeloen-korppi_stabilized_2018}, optomechanical non-reciprocity~\cite{bernier_nonreciprocal_2017, peterson_demonstration_2017}.
In addition the reported Floquet dynamics are relevant to future optomechanical experiments, such as topological photon and phonon control with dynamic modulation~\cite{schmidt2015,mathew2018,Xu_2018} and continuous variable quantum information~\cite{schmidt2012}.
Our study indicates the rich physics that remains to be explored when considering the optomechanical Hamiltonian in conjunction with other types of interactions, such as Kerr-type interactions as outlined here.

We specifically analyze this new dynamics in the context of motional sideband asymmetry~\cite{safavi-naeini_observation_2012,purdy_optomechanical_2015,underwood_measurement_2015,weinstein_observation_2014}, which has received significant attention as it is a signature of the quantum-mechanical nature of the interaction between light and engineered mechanical oscillators.
The anti-Stokes scattering (resulting in blue-shifted radiation) rate scales with $\bar n$, while the Stokes scattering rate scales with $\bar n+1$, where $\bar n$ denotes the average thermal occupation of the vibrational mode~\cite{clerk_introduction_2010}. The ratio of the two scattering rates is given by the Boltzmann factor $\exp(-\hbar\Omega_m/k_BT)$, where $\Omega_m$ is the frequency of the oscillator and $T$ its mode temperature,
which therefore allows for absolute and self-calibrated quantum noise thermometry.
Such motional sideband asymmetry has been observed in the quantized motion of laser-cooled trapped ions~\cite{diedrich_laser_1989} and cold atoms in optical lattices~\cite{jessen_observation_1992} and used for thermometry in solids~\cite{hart_temperature_1970, nemanich_raman_1984, cui_noncontact_1998} and molecules~\cite{eckbreth_laser_2000}, and it is commercially used in fiber optical distributed temperature measurements~\cite{ukil_distributed_2012}.
To exploit sideband asymmetry in optomechanical system for absolute thermometry without the need for calibration~\cite{purdy_quantum_2017}, understanding all noise contributions, such as laser noises~\cite{jayich_cryogenic_2012, sudhir_appearance_2017, kippenberg_phase_2013}, is crucial.

The novel dynamics reported here are demonstrated
by  optomechanical sideband cooling of a $5.3\unit{GHz}$ breathing mode of a nanobeam optomechanical crystal~\cite{chan_optimized_2012}, thermalized in a $^3$He buffer gas cryostat at $\sim 4.5\unit{K}$, self-calibrated at $1.5\pm 0.2$ quanta (40\% ground-state probability).
The buffer gas enables substantially improved thermalization, reducing the effect of laser-induced heating reported earlier~\cite{meenehan_silicon_2014, meenehan_pulsed_2015, riedinger_non-classical_2016,hong_hanbury_2017,riedinger_remote_2018}.
The thermal Kerr-type nonlinearity is shown to modify the motional sideband asymmetry, giving lower temperatures than the actual ones by artificially augmenting the sideband ratio.
Counterintuitively, the coupling of the sidebands due to the Kerr-type nonlinearity and the induced artificial asymmetry are even present when operating far (i.e.~$\times 10$) above the characteristic thermal response of the system. It can only be eliminated by separating the driving tones in frequency far beyond the bandwidth of the Kerr-type nonlinearity, as verified by independent measurements. The intrinsic motional sideband asymmetry is then restored, and enables performing self-calibrated thermometry.

\section{Experimental Results}
\label{sec:ExprResults}

\begin{figure*}
  \includegraphics[scale=1]{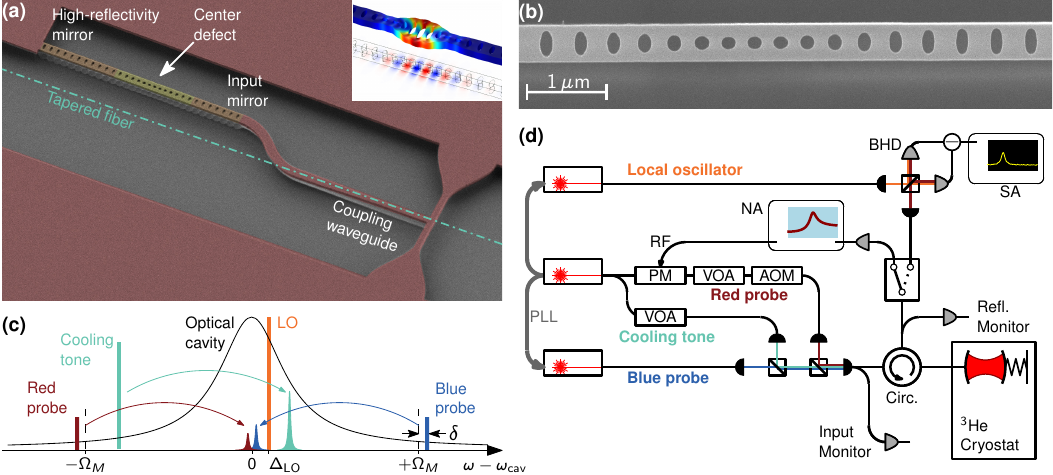}
  \caption{\textbf{Optomechanical crystal and experimental setup.}
  (a)~False-color SEM image of the silicon optomechanical crystal cavity with a waveguide for laser input coupling.
  The path of the tapered fiber is indicated.
  The inset shows the simulated mechanical breathing mode and optical mode.
  (b)~SEM image of the cavity.
  (c)~Scheme for motional sideband asymmetry measurement using two probe tones and a cooling tone.
  (d)~Experimental setup. PM, phase modulator; VOA, variable optical attenuator;
AOM, acousto-optical modulator; BHD, balanced heterodyne detector; SA, spectrum analyzer; NA, network analyzer; PLL, phase-locked loop.}
\label{fig:experimental}
\end{figure*}

Our experimental system is a nanobeam optomechanical crystal~\cite{chan_optimized_2012}, shown in Fig.~\ref{fig:experimental}.
Optically, the device functions as a single-sided cavity with a partially transmitting input mirror.
Light is evanescently coupled from a tapered optical fiber into a waveguide that forms part of the nanobeam with efficiency exceeding 50\%.
The optical resonance is at $1540\unit{nm}$ with a linewidth of $\kappa/2\pi=1.6\unit{GHz}$, of which $\kappa\s{ex} = 0.3\kappa$ are extrinsic losses to the input mirror.
The optical mode is optomechanically coupled to a mechanical breathing mode of frequency $\Omega_m/2\pi=5.3\unit{GHz}$, strongly confined due to a phononic bandgap, with an intrinsic linewidth of  $\Gamma\s{int}/2\pi=84\unit{kHz}$.
This places the system in the resolved sideband regime~\cite{schliesser_resolved-sideband_2008}, $\Omega_m>\kappa$, as required for ground-state cooling~\cite{wilson-rae_theory_2007,marquardt_quantum_2007}.
The optomechanical coupling parameter is $g_0/2\pi = 780\unit{kHz}$.
Full details of the device design, the setup and the system are given in Appendices~\ref{sec:fabrication} and~\ref{sec:expr_details}.

Though optomechanical crystals have initially been cooled to their ground state using optomechanical cooling~\cite{chan_laser_2011}, subsequent work relied on passive cooling in dilution refrigerators at $\sim 10\unit{mK}$ to reach the ground state at GHz frequencies. However, severe heating due to optical absorption precluded continuous measurements with high optomechanical cooperativities~\cite{meenehan_silicon_2014,meenehan_pulsed_2015,riedinger_non-classical_2016,hong_hanbury_2017,riedinger_remote_2018}.
Our system is passively cooled to low initial thermal occupancy at $4.5\unit{K}$ (thermal occupancy at bath temperature $\bar n\s{th}\simeq 17$) using a $^3$He buffer gas cryostat.
The buffer gas provides additional mechanical damping, increasing $\Gamma\s{int}$ by several 10s kHz to the actual mechanical linewidth $\Gamma_m$.
However,
as detailed in Appendix~\ref{sec:sideband_cooling}
and shown in the measurements described in this section, the buffer gas environment enables greatly enhanced thermalization of the oscillator and is necessary for cooling close to the ground state.

To measure motional sideband asymmetry, we employ two probe tones, around upper and lower motional sidebands, and in addition apply a strong tone near the lower sideband for sideband cooling (Fig.~\ref{fig:experimental}c).
Such multi-tone probing scheme has been applied in previous experiments in the microwave domain~\cite{weinstein_observation_2014,lecocq_quantum_2015}.
The two weak equal-power probes (about $n_c=100$ mean intracavity photons each) are applied at frequencies $\omega\s{cav}\pm(\Omega_m+\delta)$.
The cooling tone is blue-detuned from the lower probe by the frequency $\Omega\s{mod}$, where $\Gamma_m\ll\Omega\s{mod}\ll\kappa$.
The red-detuned probe will generate a resonantly enhanced anti-Stokes Raman process, where a probe photon is upconverted in frequency from $-\Omega_m-\delta$ to $-\delta$, while destroying a phonon in the mechanical oscillator.
The converse occurs for the blue-detuned probe, where a probe photon is downconverted from $\Omega_m+\delta$ to $\delta$ while creating a phonon, thus forming the resonant enhanced Stokes sideband.
The experimental setup is shown in Fig.~\ref{fig:experimental}d.
We use balanced heterodyne detection for quantum-limited measurement of the scattered thermomechanical sidebands in the output spectrum.
The overall detection  efficiency is $\eta\simeq 4\%$.
We measure the symmetrized autocorrelator of the photocurrent \cite{weinstein_observation_2014,BowenMilburn2015}
$\bar S_I(\omega)=\frac12\int_{-\infty}^\infty\langle\{\overline{\hat{I}_{\text{out}}(t+t'),\hat{I}_{\text{out}}(t')}\}\rangle e^{i\omega t}dt$, where the average over $t'$ is denoted by an overbar.
The one-sided heterodyne spectrum takes the form (in the resolved sideband limit)
\begin{equation}
  \begin{aligned}
	  {}&\bar S_I(\omega+\Delta_{\text{LO}})=1
	+\eta\Gamma_{\text{tot}}\left\{\frac{\Gamma_m\C_{\text{red}}\bar n}{\Gamma_{\text{tot}}^2/4+(\omega+\delta)^2}\right.\\
	&+\left.\frac{\Gamma_m\C_{\text{cool}}\bar n}{\Gamma_{\text{tot}}^2/4+(\omega+\delta-\Omega_{\text{mod}})^2}
	+\frac{\Gamma_m\C_{\text{blue}}(\bar n+1)}{\Gamma_{\text{tot}}^2/4+(\omega-\delta)^2}\right\},
  \end{aligned}
  \label{eq:ideal_theory}
\end{equation}
where we have introduced cooperativities $\C_{\text{red,blue,cool}}\equiv 4g_{r,b,c}^2/(\kappa\Gamma_m)$, the light-enhanced optomechanical coupling $g_{r,b,c}\equiv g_0\sqrt{n_{r,b,c}}$, reduced occupancy $\bar n=\Gamma_m\bar n\s{th}/\Gamma_{\text{tot}}$, local oscillator detuning $\Delta_{\text{LO}}=\omega_{\text{LO}}-\omega_{\text{cav}}$, and incorporated the effect of sideband cooling into a broadened mechanical linewidth $\Gamma_{\text{tot}} \simeq \Gamma_m (1+\C_{\text{cool}})$.
In the last expression we assume weak probe tones of equal strength, such that they do not contribute to the total mechanical damping, and taken the good cavity limit $\kappa/\Omega_m\to\infty$, thus neglecting the quantum backaction limit to optomechanical cooling~\cite{wilson-rae_theory_2007,marquardt_quantum_2007}, which is of no importance here.
We further assume that the cavity linewidth is much larger than the detuning and effective mechanical linewidth $\kappa\gg \delta,\Gamma_{\text{tot}}$, such that the optical susceptibility can be evaluated on resonance, and neglect classical laser noise.
Equation~\eqref{eq:ideal_theory} is normalized to the shot noise floor, includes the overall detection efficiency $\eta$,
and we have chosen to show only sidebands close to resonance, the others are heavily suppressed due to the cavity resonance.

For our measurement, the red and blue cooperativities are chosen to be equal $\C_{\text{red}}=\C_{\text{blue}}\equiv\C$,
such that the Lorentzian probe tone sidebands centered around $\Delta_{\text{LO}}\pm\delta$
have weights proportional to $\C \bar n$ and $\C(\bar n+1)$.
The asymmetry can be interpreted as a consequence of either quantum backaction (i.e., the optical input vacuum fluctuations) or of the mechanical commutation relations,
depending on the detection scheme as well as the detector model used \cite{khalili_quantum_2012,weinstein_observation_2014,borkje_heterodyne_2016}.
From the sidebands we extract $\bar n$ either directly from the red-detuned probe sideband weight (with suitable calibration as outlined in Appendix~\ref{sec:sideband_cooling})
or from the ratio of the weights of the two sidebands, as in previous optomechanical experiments~\cite{purdy_optomechanical_2015, underwood_measurement_2015}.
Raman thermometry can be done using the cooling tone sideband instead of the red-detuned probe sideband, as long as one accounts for the difference in optomechanical coupling.
Classical laser noise, which can be a source of artificially enhanced asymmetry~\cite{jayich_cryogenic_2012,safavi-naeini_laser_2013,sudhir_appearance_2017} is not affecting our measurements, as is shown in Appendix~\ref{sec:excess_noise}.

\begin{figure*}[ht]
  \includegraphics[scale=1]{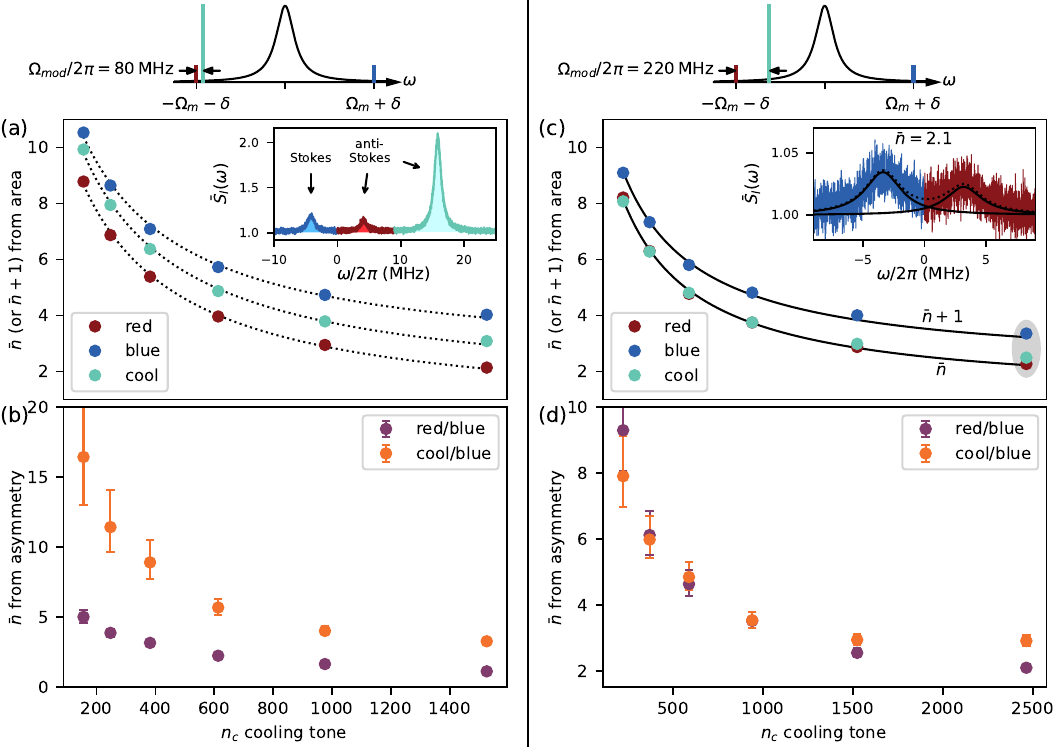}
  \caption{\textbf{Artificial and quantum sideband asymmetry in optomechanical sideband cooling.} In (a), (b), the cooling tone is detuned $80\unit{MHz}$ from red-detuned probe.
  (a)~Inferred $\bar n$ ($\bar n+1$) from anti-Stokes (Stokes) mechanical sidebands, based on total thermomechanical noise.
  Overall cooling is observed, however a clear discrepancy exists since the cooling tone and the red-detuned probe should yield the same and be equal to the occupancy $\bar n$.
  Inset: Example of an actual observed noise spectrum, colored for different tones as in the main panel.
  Note that the choice of local oscillator frequency (Fig.~\ref{fig:experimental}c) leads to the noise spectrum `folded over' and to the cooling tone sideband located close to the probe sidebands.
  The $\omega$-axis does not strictly correspond to Eq.~\eqref{eq:ideal_theory} in this case.
  (b)~The occupancy $\bar n$ obtained from motional sideband asymmetry, obtained both from the ratio of red- and blue-detuned probes, and the ratio of cooling tone and blue-detuned probe, which shows strong disagreement.
  Error bars represent $\pm 20\unit{MHz}$ tuning accuracy, see text for discussion.
  (c), (d) show the data corresponding to (a), (b), respectively, only with the cooling tone detuned $220\unit{MHz}$ from red-detuned probe, where the effect of the thermal Kerr-type nonlinearity is strongly diminished.
  In~(c), the lower curve is a fit according to the model of Appendix~\ref{sec:sideband_cooling} with the average asymmetry of the last two points, shown in (d), used for calibration; one quantum is added to result in the upper curve, coinciding with the Stokes sideband data.
  Inset (c)~shows spectra of the last data point.
}
\label{fig:fake_asymmetry}
\end{figure*}

Throughout the experiment, the power of the cooling tone is increased, lowering the mechanical occupancy $\bar{n}$ through sideband cooling, while the probe tones are held constant.
Figure~\ref{fig:fake_asymmetry}a,b we show thermometry results for a detuning of the red-probe from the cooling tone of $\Omega\s{mod}/2\pi=80\unit{MHz}$.
In Fig.~\ref{fig:fake_asymmetry}a we use the total sideband power to infer $\bar n$ ($\bar n+1$) from the anti-Stokes (Stokes) mechanical sidebands, plotted against the cooling tone intracavity photons.
Strikingly, the curves of the cooling tone and red-detuned probe \emph{do not coincide}, making it impossible to associate $\bar n$ with either.
The discrepancy is also reflected in Fig.~\ref{fig:fake_asymmetry}b, where  \emph{quantum sideband asymmetries} of either of the two anti-Stokes sidebands compared to the Stokes sideband yield \emph{different} $\bar n$.
We next repeat the measurement with larger seperation between the red-detuned probe and the cooling tone, with
$\Omega\s{mod}/2\pi=220\unit{MHz}$ (Fig.~\ref{fig:fake_asymmetry}c,d).
The inferred $\bar n$ from both anti-Stokes sidebands \emph{now show excellent agreement} (Fig.~\ref{fig:fake_asymmetry}c), and $\bar n$ inferred from motional sideband asymmetry also agree within experimental errors
(Fig.~\ref{fig:fake_asymmetry}d). Our measurements thus show that the presence of the cooling tone, when tuned closely to the red-detuned probe, modifies the quantum-mechanical motional sideband asymmetry.

\begin{figure*}
  \includegraphics[width=\linewidth]{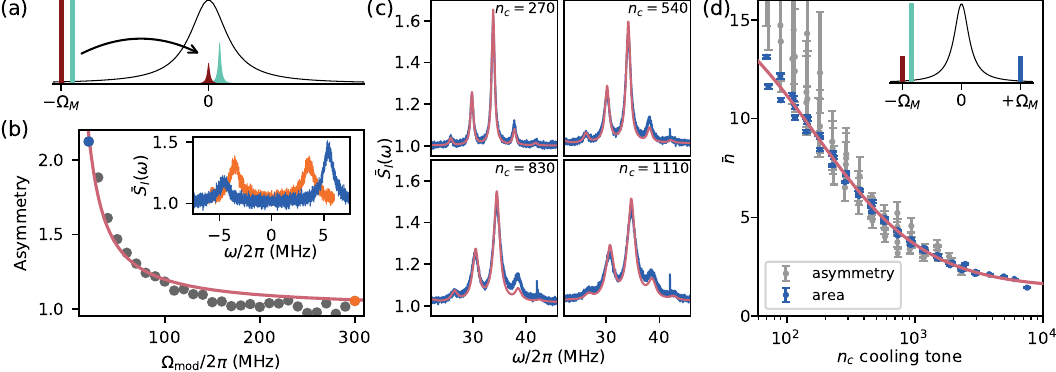}
  \caption{\textbf{Observation of asymmetric noise spectra due to Kerr-type nonlinearity.}
  (a)~Pumping scheme with two equal red-detuned pumps.
  (b)~The peak ratio of the observed spectra, relative to the ratio expected from bare optomechanical response, vs.~$\Omega\s{mod}$ for constant power of $n\s{c}=640$ intracavity photons, showing decreased effect for higher modulation frequencies.
  The inset shows the spectra for data points of the same color in the main panel.
  (c)~Increasing the pump power for a constant $\Omega\s{mod}/2\pi=4\unit{MHz}$, showing  higher-order Fourier modes.
  Also shown are fits to the analytic model of Sec.~\ref{sec:theory}, see text for details.
  (d)~Calibration of sideband cooling using motional sideband asymmetry, with $\Omega\s{mod}/2\pi=220\unit{MHz}$, free from Kerr-type artificial asymmetry. See text for more details. The oscillator is cooled down to $\bar n=1.5$ phonons.
}
\label{fig:dual_tone_all}
\end{figure*}

To investigate this phenomenon further, we perform an auxiliary experiment, shown in Fig.~\ref{fig:dual_tone_all}a.
Omitting the blue-detuned probe, we apply two \emph{equal power} tones near the lower mechanical sideband (still referred to as red-detuned probe and cooling tone).
As illustrated in Fig.~\ref{fig:dual_tone_all}a, the observed anti-Stokes sidebands are not equal even for $\Omega\s{mod}\ll\kappa$, with the higher-frequency sideband stronger, in disagreement with standard optomechanical theory.
Keeping the red-detuned probe fixed at $\Delta=-\Omega_m$ and scanning the cooling tone, Fig.~\ref{fig:dual_tone_all}b shows the ratio of the two sidebands, normalized to the expected bare optomechanical response, vs.~$\Omega\s{mod}$.
The normalized ratio decreases with $\Omega\s{mod}$, only reaching the expected behavior at $\Omega\s{mod}\gtrsim 200\unit{MHz}$.
Note that, depending on the power used, the ratio may be very large for small $\Omega\s{mod}$, e.g.~exceeding 2 for $\Omega\s{mod}/2\pi=20\unit{MHz}$ in Fig.~\ref{fig:dual_tone_all}b.
Figure~\ref{fig:dual_tone_all}c shows the noise spectra at low $\Omega\s{mod}/2\pi=4\unit{MHz}$ for increasing tone power, where additional higher-order sidebands, spaced by $\Omega\s{mod}$, are observed.
These measurements point to the presence of a cavity nonlinearity, which couples the thermomechanical sidebands and modifies the observed asymmetry.

The existence of cavity nonlinearities due to e.g.~thermal effects is not unusual and has been observed in both bulk optical cavities as well as ultrasmall optical mode volume resonators~\cite{ilchenko_thermal_1992,fomin_nonstationary_2005,rokhsari_observation_2005,barclay_nonlinear_2005}.
While there are several potential sources for a ``Kerr-type'' effect, we consider the photothermorefractive frequency shift (PTRS) as the dominant mechanism.
Physically, photons circulating in the cavity are absorbed, leading to heating, and thus shifting of the cavity resonance (e.g., via the temperature dependent refractive index). The temperature deviation $\delta T$ is governed in the simplest case by a single timescale and temperature-independent absorption
\begin{align}
	\delta\dot T(t)&=-\gamma\s{th}\delta T(t)+g_{\text{abs}}|\bar a(t)|^2.
	\label{eq:temperature_deviation}
\end{align}
Here the absorption rate is given by $g_\mathrm{abs}$, and the thermalization rate by $\gamma_\mathrm{th}$.
In the presence of two tones, the cavity field intensity beats $n_c(t)=|\bar a(t)|^2\propto\text{const}+\cos(\Omega_{\text{mod}}t)$, which, via the absorption heating, causes a periodic modulation of the cavity frequency, captured by the detuning
\begin{equation}
	\Delta\s{th}(t)=\Delta_k \exp(i\Omega_{\text{mod}}t)+\text{c.c.}
	\label{eq:PTRS}
\end{equation}
where
\begin{equation}
	\Delta_{k}=\frac{g\s{PT}g_{\text{abs}}\bar a_{c}\bar a_{r}}{\sqrt{\gamma \s{th}^2+\Omega_{\text{mod}}^2}}e^{-i\phi_{\text{th}}},
	\quad \phi_{\text{th}}=\tan^{-1}\left(\frac{\Omega_{\text{mod}}}{\gamma_{\text{th}}}\right).
	 \label{eq:Delta_k}
\end{equation}
Here $\bar a_{c},\bar a_{r}$ are the amplitudes of the intracavity fields produced by cooling tone and red-detuned probe, and
$g_{\text{PT}}$ relates cavity frequency shift to temperature deviation via $\Delta\s{th}=g\s{PT}\delta T$. Consequently the static thermal cavity shift per mean intracavity photon is given by $g\s{0,th}=g\s{PT}{g\s{abs}}/\gamma\s{th}$.
In principle, the nonlinear cavity frequency shift could have a number of origins, which can be included in $\Delta_k$. Here for simplicity we consider only the PTRS \eqref{eq:PTRS}, such that $\Delta_k$ is given by Eq.~\eqref{eq:Delta_k}.
We consider additional Kerr-type nonlinearities in the end of Sec.~\ref{sec:theory}.

The cavity frequency modulation mediates processes in which photons are scattered from a frequency $\omega$ to $\omega\pm\Omega_{\text{mod}}$, which causes coupling of the sidebands with strength $\Delta_k$.
In Section~\ref{sec:theory}, we incorporate the PTRS into standard optomechanical theory to model our experiments, and use a Floquet approach, based on writing the cavity and mechanical annihilation operators as sums of Fourier modes, to solve the time-dependent quantum Langevin equations.

In the first approximation, corresponding to retaining the dominant Fourier modes, the output spectrum takes the same form as above \eqref{eq:ideal_theory} but with modified cooperativities
\begin{subequations}
  	\begin{align}
  		\tilde\C\s{red}&=\C\s{red}\left|1-2ig_c\Delta_k /(g_r\kappa)\right|^2,\\
		\tilde\C\s{cool}&=\C\s{cool}\left|1-2ig_r\Delta_k ^*/(g_c\kappa)\right|^2,
  	\end{align}
  	\label{eq:modified_cooperativities}
\end{subequations}
leaving the ideal theory [Eq.~\eqref{eq:ideal_theory}] otherwise unchanged.
Importantly, these expressions can lead to an asymmetry when $\Delta_k$ is complex even when $g_r=g_c$.
This explains our observations in Fig.~\ref{fig:dual_tone_all}b, where the cavity is driven by two equal-strength pumps, and asymmetric sidebands are observed.
The asymmetry diminishes as $\Omega\s{mod}$ is increased beyond the bandwidth $\gamma\s{th}$.
A fit to this simple model shown in Fig.~\ref{fig:dual_tone_all}b is in good agreement and captures this general behavior.
In this case of $\Omega\s{mod}\gg\gamma\s{th}$ we find that the approximation~\eqref{eq:modified_cooperativities} is sufficient to describe our data, with added Fourier modes producing no change to the  fitted curve.
From the fit we obtain the two quantities characterizing $\Delta_k$~\eqref{eq:Delta_k}, $\gamma\s{th}/2\pi\sim 6\unit{MHz}$ and $g\s{abs}g\s{PT}/4\pi^2\sim 10\unit{MHz^2}$, with other parameters determined independently.
Deviations from the data at high frequencies may be accounted for by considering additional Kerr-type effects of higher bandwidths or more complicated thermal behavior than that afforded by Eq.~\eqref{eq:temperature_deviation}, however we keep the simplicity of our model and emphasize that for $\Omega\s{mod}/2\pi=220\unit{MHz}$ used in Fig.~\ref{fig:fake_asymmetry}c,d and all our subsequent motional sideband asymmetry measurements, we consistently observe the same normalized sideband power (i.e.~same $\bar n$) for both cooling tone and red-detuned probe (within experimental errors of $\sim 5\%$), confirming the weakness of Kerr-type effects, and allowing to operate in a regime where the quantum sideband asymmetry arising from optomechancial quantum correlations can be observed.
We also note that the coupling of red and blue tones is negligible as their spacing is $2\Omega_m\gg\gamma\s{th}$.

In order to capture the Floquet dynamics observed in Fig.~\ref{fig:dual_tone_all}c, where a series of thermomechanical sidebands are generated, more Fourier modes have to be included (six optical and five mechanical; see Fig.~\ref{fig:fourier_modes}).
The relative weight of the higher order sidebands is in good agreement with the data.
Strong thermal effects at $\Omega\s{mod}/2\pi=4\unit{MHz}$ lead to distortion in the cavity linewidth measurement, introducing large detuning errors, leading to large uncertanties in cavity parameters.
The fits shown in Fig.~\ref{fig:dual_tone_all}c yield $\gamma\s{th}/2\pi$ in the range 4--$8\unit{MHz}$ and $g\s{abs}g\s{PT}/4\pi^2\sim 6$--$17\unit{MHz}^2$, in close agreement with Fig.~\ref{fig:dual_tone_all}b.
In order to further confirm our model, we have carried out pump-probe response measurements of the cavity, detailed in Appendix~\ref{sec:frequency_response}.
We found the bandwidth of power induced cavity frequency shifts (likely thermal in origin) $\sim 10\unit{MHz}$, in agreement with the results presented here.

Next, we turn back to sideband measurements at $\Omega\s{mod}/2\pi=220\unit{MHz}$, where only the quantum asymmetry is prominent.
Figure~\ref{fig:dual_tone_all}d shows an extended set of measurements done at cryostat temperature of $4.4\unit{K}$ and buffer gas pressure of $140\unit{mbar}$, including occupancies $\bar n$ inferred from both motional sideband asymmetry and power in the cooling sideband.
Inferring $\bar n$ from motional sideband asymmetry is straightforward, however the probes must be weak to avoid extraneous heating of the oscillator.
The much-larger signal-to-noise ratio of the cooling sideband is more suitable for determination of $\bar n$ for the highest cooling powers.
Referring to Eq.~\eqref{eq:ideal_theory} and neglecting the weak probes, we see that apart from the easily-measured optomehcanical parameters, accurate knowledge of the quantity $\eta\,\bar n\s{th}$ is required.
Moreover, extraneous heating due to optical absorption modifies the actual bath occupancy from that measured, $\bar n\s{th}\rightarrow\bar n\s{th}+\heating n_c$, where $\heating$ is the added bath phonons per intracavity photon (see also Appendix~\ref{sec:sideband_cooling}).
Thus, at least one of the parameters $\eta,\bar n\s{th},\heating$ needs to be independently determined.
It is often difficult to obtain accurate measurement of these parameters.
Here we use the sideband asymmetry at an intermediate data point to compute $\bar n$ and, unequivocally, $\eta$, providing calibration for the entire cooling curve of Fig.~\ref{fig:dual_tone_all}d.

The two main sources of measurement error are tuning accuracy, estimated at
$\pm 2\pi\times 20\unit{MHz} = \pm 0.0125\kappa$, that leads to slightly different cavity response seen by the two probes (error bars in Figs.~\ref{fig:fake_asymmetry} and~\ref{fig:dual_tone_all}d), and error in estimate of Lorentzian peaks of the sidebands, that yield different $\bar n$ values for the two flavors of asymmetry used in Fig.~\ref{fig:fake_asymmetry}b,d (see inset of Fig.~\ref{fig:fake_asymmetry}c).
The former error is dominant for large $\bar n$ (small asymmetry), while the latter is dominant at strong cooling powers hence low signal-to-noise ratios (for example the last point in Fig.~\ref{fig:fake_asymmetry}d).
For every data point in Fig.~\ref{fig:dual_tone_all}d, we add these errors in quadrature.
We then compute $\eta$ and its uncertainty using weighted average.
The final result, $\eta = 0.044\pm 0.005$ remains essentially the same if we take the last few, minimum error points.
This calibration gives $\bar{n}=1.5\pm 0.2$ (40\% ground-state occupation) for the minimum occupation in Fig.~\ref{fig:dual_tone_all}d.
In addition, fitting the entire data set yields bath thermal occupation $\bar n\s{th}=17.5$ and extraneous heating of $\heating=1.3\,\C_0$, in excellent agreement with the independently calibrated measurements of Appendix~\ref{sec:sideband_cooling}.

\section{Theory
}
\label{sec:theory}

To describe our experiment we take the standard optomechanical model in a rotating frame
\begin{equation}
  	\hat{H}\s{OM}/\hbar = [\Delta_{\text{th}}(t)-\Delta]\hat a\dagg\hat a
  + \Omega_m\hat b\dagg\hat b
  - g_0\hat a\dagg\hat a(\hat b+\hat b\dagg),
  \label{eq:HOM}
\end{equation}
but include the PTRS (via $\Delta_{\text{th}}$) as well as optical and mechanical baths.
A standard calculation~\cite{gardiner2004quantum,aspelmeyer_cavity_2014} gives quantum Langevin equations,
which we linearize around the mean intracavity field $\hat a(t)=\bar a(t)+\delta\hat a(t)$ (and the same for the mechanical mode) to obtain the equations for linear optomechanics~\cite{aspelmeyer_cavity_2014}
\begin{equation}
  \begin{aligned}
	\delta \dot{\hat a} &= \left\{i\left[\Delta-\Delta\s{th}(t)\right]-\frac{\kappa}{2}\right\}\delta\hat a
	+ig(t)(\delta\hat b+\delta\hat b\dagg)+\sqrt{\kappa}\delta\hat a\s{in},\\
	\delta\dot{\hat b} &= \left(-i\Omega_m-\frac{\Gamma_m}{2} \right)\delta\hat b
	+i[g(t)\delta\hat a\dagg+g^*(t)\delta\hat a]+\sqrt{\Gamma_m}\hat b\s{in},
  \end{aligned}
  \label{eq:langevin_eoms}
\end{equation}
where $\Delta\approx-\Omega_m$ is the average detuning of the pumps from the cavity,
$g(t)=g_0\bar a(t)\exp[i(\omega\s{cav}+\Delta)t]$ is the modulated light-enhanced optomechanical coupling strength,
and the input noises obey $\langle\delta\hat a\s{in}(t)\delta\hat a\s{in}\dagg(t')\rangle=\delta(t-t')$ and $\langle\hat b\s{in}(t)\hat b\s{in}\dagg(t')\rangle=(\bar n\s{th}+1)\delta(t-t')$ (as well as standard commutation relations).
Langevin equations with periodic time-dependence can be analyzed with a recently developed method~\cite{malz_floquet_2016}.
Note that closely related models have been studied in the context of levitated optomechanics~\cite{Aranas2016,Aranas2017}, where instead of the cavity, the mechanical resonator frequency is modulated.

In the experiment, we apply up to three tones to the cavity: a strong cooling tone, as well as weak red- and blue-detuned probes. Since the blue-detuned probe is very far detuned from the other two tones, it remains unaffected and we will not consider it in this section.
Cooling tone and red-detuned probe are close to the red sideband, separated in frequency by $\Omega_{\text{mod}}$, as is shown in Fig.~\ref{fig:dual_tone_all}a.
Neglecting all other effects, the resulting intracavity field $\bar a(t)=\bar a_{r}e^{-i(\omega_{\text{cav}}+\Delta)t}+\bar a_{c}e^{-i(\omega_{\text{cav}}+\Delta+\Omega_{\text{mod}})t}$.

We model the PTRS~\cite{verhagen_quantum-coherent_2012,Li2014} through Eq.~\eqref{eq:temperature_deviation} in conjunction with $\Delta\s{th}(t)=g\s{PT}\delta T(t)$ as discussed above.
In our level of approximation we neglect the effect of the PTRS on the mean intracavity field, such that we can solve directly for the temperature deviation.
Since a constant temperature shift leads to a static frequency shift which can be absorbed into the overall detuning $\Delta$,
we only consider the time-dependent part
\begin{equation}
  \begin{aligned}
	\delta T(t)&=g_{\text{abs}}\int_{-\infty}^t2\bar a_{c}\bar a_{r}e^{-\gamma\s{th}(t-t')}\cos(\Omega_{\text{mod}}t')dt'\\
	&=\frac{2g_{\text{abs}}\bar a_{c}\bar a_{r}}{\sqrt{\gamma\s{th}^2+\Omega_{\text{mod}}^2}}\cos(\Omega_{\text{mod}}t-\phi\s{th}),
  \end{aligned}
  \label{eq:thermal_response}
\end{equation}
where the phase lag
\begin{equation}
  \phi\s{th}=\tan^{-1}\left(\Omega\s{mod}/\gamma\s{th}\right)
  \label{eq:phase_lag}
\end{equation}
arises due to the finite thermalization time.
Equation \eqref{eq:thermal_response} yields the PTRS displayed in Eqs.~\eqref{eq:PTRS} and \eqref{eq:Delta_k}.
The phase $\phi\s{th}$ plays a crucial role in the observed asymmetric response. Note that if $\Delta\s{th}$ is positive, the cavity resonance is blue-shifted.
Absorption of photons in the cavity can also lead to a delayed photothermal force on the mechanical oscillator due to the photo-thermo-elastic effect~\cite{DeLiberato2011,Restrepo2011,Metzger2004,shkarin2019}, which in turn leads to a cavity frequency shift via the optomechanical coupling.
In our experiment the photothermal body force is negligible due to the high mechanical frequency relative to the thermal bandwidth, thus we focus on the photo-thermo-refractive effect.

\begin{figure}
  \includegraphics[width=\linewidth]{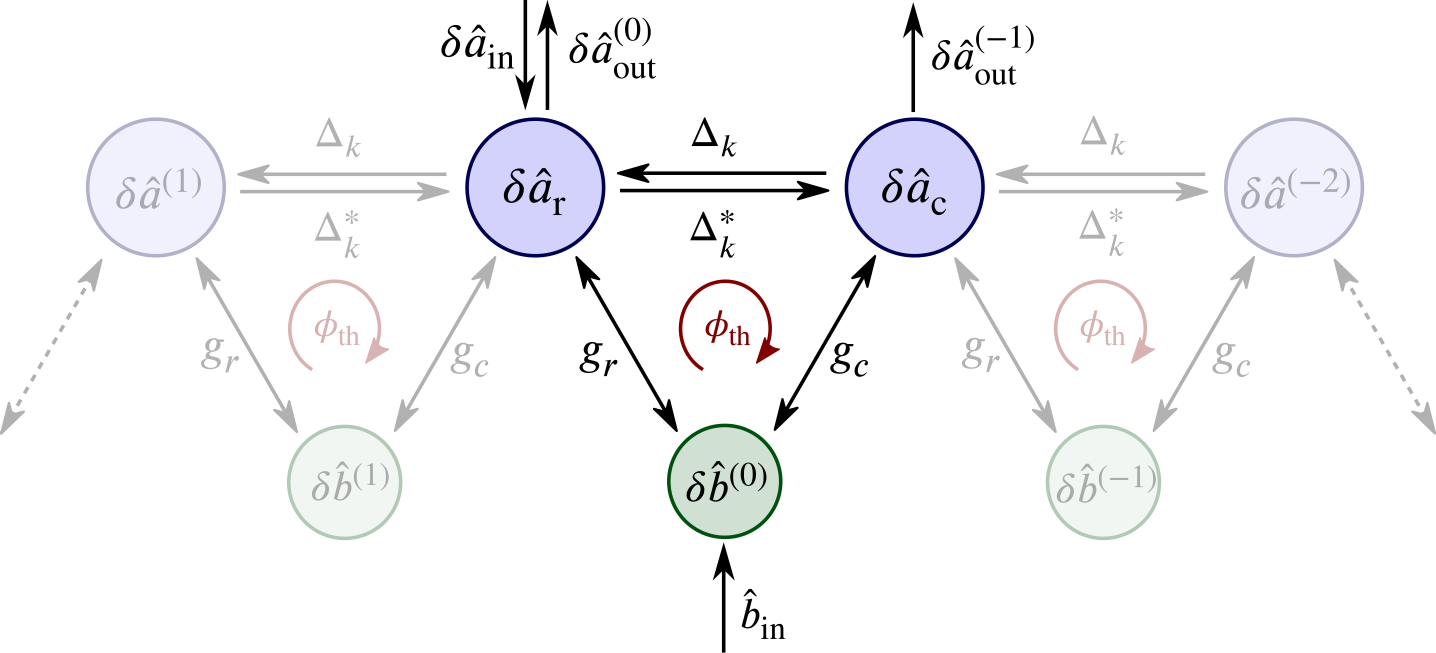}
  \caption{\textbf{An illustration of the infinite array of coupled Fourier modes.}
  The map to Fourier modes \eqref{eq:fourier_cpts} introduces an infinite set of coupled Langevin equations, which has to be truncated at some order to obtain a solution.
  Due to the two pumps (cooling and red-detuned probe), thermomechanical noise incident on mode $\delta\hat b^{(0)}$ is distributed onto $\delta\hat a_r$ and $\delta\hat a_{c}$, generating two sidebands.
  The cavity Fourier modes are coupled due to a cavity Kerr-type nonlinearity [$\Delta_k $~\eqref{eq:Delta_k}].
  This effect modifies the sideband weight, which to lowest order (considering only the bold three-mode system)
  can be accounted for with new effective optomechanical couplings~\eqref{eq:modified_cooperativities},
  but in general leads to more sidebands (Fig.~\ref{fig:dual_tone_all}c), modeled by including more Fourier modes in the description.
  Notably, due to the finite response time, the coupling between the Fourier modes is \emph{complex}, with the phase $\phi_{\text{th}}$ given in Eq.~\eqref{eq:phase_lag}.}
  \label{fig:fourier_modes}
\end{figure}

If the two pumps lie close to the red sideband ($\Delta\approx-\Omega_m$), and the modulation frequency is much smaller than $\Omega_m$, as is the case here, we can neglect terms rotating at $2\Omega_m$ in a rotating-wave approximation.
However, the resulting equations still have an explicit time-dependence, which can be removed by splitting the fields $\delta\hat a$ and $\delta\hat b$ into Fourier modes~\cite{malz_floquet_2016}
\begin{equation}
  \begin{aligned}
    \delta\hat a(t)&=\sum_n\exp(in\Omega\s{mod}t)\delta\hat a^{(n)}(t),\\
    \delta\hat b(t)&=\sum_n\exp(in\Omega\s{mod}t)\delta\hat b^{(n)}(t),
  \end{aligned}
  \label{eq:fourier_cpts}
\end{equation}
at the cost of introducing an infinite set of coupled equations of motion
\begin{equation}
  \begin{aligned}
	\delta \dot{\hat a}^{(n)}&=\left( i\tilde\Delta-in\Omega_{\text{mod}}-\frac{\kappa}{2} \right)\delta\hat a^{(n)}+\delta_{n,0}\sqrt{\kappa}\delta\hat a_{\text{in}}\\
	&+i\left(g_{c} \delta\hat b^{(n+1)}+g_{r}\delta\hat b^{(n)}-\Delta_k ^*\delta\hat a^{(n+1)}-\Delta_k \delta\hat a^{(n-1)} \right),\\
	\delta\dot{\hat b}^{(n)}&=\left(-in\Omega_{\text{mod}}-\frac{\Gamma_m}{2}\right)\delta\hat b^{(n)}\\
	&+i\left( g_{c}\delta\hat a^{(n-1)}+g_{r}\delta\hat a^{(n)}\right)+\delta_{n,0}\sqrt{\Gamma_m}\hat b_{\text{in}},
  \end{aligned}
  \label{eq:eom}
\end{equation}
which are depicted as lattice in Fig.~\ref{fig:fourier_modes}.
Here, $g_{r,c}=g_0\bar a_{r,c}$ and we have written $\Delta\s{th}(t)=\Delta_k \exp(i\Omega_{\text{mod}}t)+\text{c.c.}$, and defined the residual detuning $\tilde\Delta=\Omega_m+\Delta\ll\kappa$, which presents only a small correction and will thus be neglected in the following.

The Fourier modes we have introduced are sometimes called auxiliary modes~\cite{peterson_demonstration_2017}, frequency-shifted operators \cite{Aranas2016}, or sidebands.
The explicitly time-dependent terms in the linearized equations of motion~\eqref{eq:langevin_eoms} couple the Fourier modes.
This off-diagonal coupling (in Fourier space) is suppressed by the response of the modes (especially the narrow mechanical mode), such that good approximations are obtained with only few Fourier modes.

For now, we assume that only the central modes $\delta\hat a_r\equiv\delta\hat a^{(0)}$, $\delta\hat a_{c}\equiv\delta\hat a^{(-1)},\delta\hat b^{(0)}$ are nonzero.
These modes are shown in bold colors in Fig.~\ref{fig:fourier_modes} and are most relevant as they contain the most thermomechanical noise.
This reduces the equations of motion~\eqref{eq:eom} to the matrix equation
\begin{equation}
  \begin{aligned}
	&\mat{\chi\s{opt}^{-1}(\omega)&-ig_{r}&i\Delta_k \\
	-ig_{r}&\chi_m^{-1}(\omega)&-ig_{c}\\
	i\Delta_k ^*&-ig_{c}&\chi\s{opt}^{-1}(\omega+\Omega\s{mod})}
	\mat{\delta\hat a_r\\\delta\hat b^{(0)}\\ \delta\hat a_{c}}\\
	&=\mat{\sqrt{\kappa}\delta\hat a\s{in}\\ \sqrt{\Gamma_m}\hat b\s{in}\\ 0},
  \end{aligned}
  \label{eq:matrix_equation}
\end{equation}
where we have defined the susceptibilities
\begin{equation}
  \chi\s{opt}^{-1}(\omega)=\kappa/2-i(\omega+\tilde\Delta), \quad\chi_m^{-1}(\omega)=\Gamma_m/2-i\omega.
  \label{eq:susceptibilities}
\end{equation}
Note that the Kerr-type effect thus leads to a coupling of the Fourier modes $\delta\hat a_{c}$ and $\delta\hat a_r$, captured by $\Delta_k $.
The noise spectrum contains mostly noise from the mechanical oscillators, such that we can neglect $\delta\hat a_{\text{in}}$.
Applying the red-detuned probe and the cooling tone introduces optical damping, yielding an effective susceptibility $\chi\s{m,eff}^{-1}(\omega)=\Gamma_m(1+\C_{\text{red}}+\C_{\text{cool}})/2-i\omega$.
Given the approximation \eqref{eq:matrix_equation}, the spectrum reads

\begin{equation}
  \begin{aligned}
  	  \bar S_I&(\omega+\Delta_{\text{LO}})=1+\eta\kappa\bar n\Gamma_m
  	  \left|\frac{g_r-ig_c\Delta_k\chi\s{opt}(\Omega\s{mod}+\omega)}{\chi\s{opt}^{-1}(\omega)\chi\s{m,eff}^{-1}(\omega)}\right|^2\\
  	  &+\eta\kappa\bar n\Gamma_m\left|\frac{g_c-ig_r\Delta_k^*\chi\s{opt}(\omega-\Omega\s{mod})}{\chi\s{opt}^{-1}(\omega)\chi\s{m,eff}^{-1}(\omega-\Omega\s{mod})}\right|^2
  \end{aligned}
  \label{eq:three_mode_spectrum}
\end{equation}
where we have neglected the frequency dependence of the cavity response,
and introduced the overall detection efficiency $\eta$~\cite{gardiner2004quantum,BowenMilburn2015}.
This spectrum can be understood as follows:
thermomechanical noise is filtered by the mechanical response $\chi_{\text{m,eff}}$ and is scattered to the optical modes, where it interferes with itself. In the Fourier mode corresponding to the red-detuned probe sideband, the amplitudes $g_r$ and $ig_c\Delta_k \chi\s{opt}(\Omega\s{mod})$ add.
The rate $g_{r}$ comes from scattering directly into that Fourier mode, whereas $ig_{c}\Delta_k \chi\s{opt}(\Omega\s{mod})$ has the clear interpretation of noise scattering first into cooling mode $\delta\hat a_{c}$ with amplitude $g_c$, where it picks up the optical susceptibility $\chi\s{opt}(\Omega\s{mod})$, and then hopping from there onto the red-detuned probe mode with amplitude $\Delta_k$.

We find that even for equal pump strength ($g_{r}=g_{c}$) the thermomechanical sideband weights can differ, \emph{as long as $\Delta_k $ has a nonzero phase}.
This phase occurs only if $\Omega\s{mod}$ and $\gamma\s{th}$ are comparable, in which case the phase lag of the thermal response behind the intracavity field [Eq.~\eqref{eq:PTRS}] is non-trivial.
This phase is conjugated between the clockwise and counterclockwise process, allowing us to interpret it as a synthetic gauge field $\phi_{\text{th}}$ threading the triangles in Fig.~\ref{fig:fourier_modes}.
From this point of view, the interpretation is similar to the non-reciprocal noise scattering observed in the optomechanical isolator~\cite{bernier_nonreciprocal_2017}, and the optoelectromechanical transducer~\cite{moaddel_haghighi_sensitivity-bandwidth_2018}, except that here the phase arises dynamically, among the virtual Fourier modes, whereas in these examples the phase is a direct consequence of the phase relation of drives.
In the more relevant case for thermometry, $g_{r}\ll g_{c}$, we can neglect the backaction of the probe tones and Eq.~\eqref{eq:three_mode_spectrum} reverts to the ideal theory \eqref{eq:ideal_theory}, but now with the modified optomechanical cooperativities displayed above [Eq.~\eqref{eq:modified_cooperativities}].
The three-mode approximation is sufficient for fitting our data in Fig.~\ref{fig:dual_tone_all}b, where mostly $\Omega\s{mod}\gg\gamma\s{th}$.

For strong driving and small $\Omega\s{mod}\sim\gamma\s{th}$, the interaction $\Delta_k $ between the optical Fourier modes is enhanced [Eq.~\eqref{eq:Delta_k}],
and thus higher-order Fourier modes are populated, and more sidebands appear in the output spectrum.
This is precisely what we observe in Fig.~\ref{fig:dual_tone_all}c.
The higher-order sidebands can be modeled by including more Fourier modes (faint color in Fig.~\ref{fig:fourier_modes}).
The matrix in Eq.~\eqref{eq:matrix_equation} is straightforwardly generalized to larger systems.
It can then be shown~\cite{malz_floquet_2016} that the normal-ordered time-averaged noise spectrum is given by
\begin{equation}
  S^N(\omega)=\sum_n\int\frac{d\omega'}{2\pi}\langle \delta\hat a\s{out}^{(n) \dagger}
  (\omega+n\Omega\s{mod})\delta\hat a\s{out}^{(-n)}(\omega')\rangle.
  \label{eq:measured_spectrum}
\end{equation}
The heterodyne spectrum we quote above is related through $\bar S_I(\omega+\Delta_{\text{LO}})=1+\eta S^N(\omega)$, which is shown explicitly in Appendix~\ref{app:heterodyne_spectrum}.
In Fig.~\ref{fig:dual_tone_all}c, to capture the full Floquet dynamics, we fit the data to this model including 6~optical Fourier modes and 5~mechanical Fourier modes
[i.e.~$d^{(2)}\ldots d^{(-3)}$, $b^{(2)}\ldots b^{(-2)}$ in Fig.~\ref{fig:fourier_modes}].

We note that the intrinsic Kerr effect also leads to a cavity frequency shift $\Delta\s{Kerr}(t) = g\s{Kerr}|\bar a(t)|^2$ without a phase lag.
The coupling strength $g\s{Kerr}$ can be estimated through~\cite{Matsko2005}
\begin{equation}
  g\s{Kerr} = -\omega\s{cav}\frac{n_2}{n_0}\frac{\hbar\omega\s{cav}c}{V\!\s{mode}n_0},
\end{equation}
where $n_0$ is the linear refractive index, $n_2$ the Kerr coefficient, $V\!\s{mode}$ the mode volume, and $c$ the speed of light.
For our system we find $g\s{Kerr}/2\pi\sim -13\unit{kHz}$, considerably weaker than the PTRS which has a static, single intracavity photon coupling $g\s{0,th}/2\pi=1.7\unit{MHz}$.
Additionally, the optomechanical interaction itself gives rise to an instantaneous Kerr-type frequency shift,
$\Delta\s{OM}(t)=g\s{OM}|\bar a(t)|^2$ with $g\s{OM}=-2g_0^2/\Omega_m\simeq- 2\pi \times230\unit{Hz}$~\cite{aspelmeyer_cavity_2014}, negligible in our system compared to the PTRS as well.
The optomechanical Kerr effect, however, can be strong in other systems~\cite{gupta2007,leijssen2017}.
The intrinsic and optomechanical Kerr effects are instantaneous and thus independent of pump spacing.

\section{Conclusion}
\label{sec:conclusion}

Rapid advances in cavity optomechanics over the last decade now enable quantum control of mechanical oscillators using electromagnetic radiation.
Here we have shown that quantum effects such as motional sideband asymmetry can be masked by classical nonlinearities, that lead to modification of the effective scattering rates between thermomechanical sidebands of closely-tuned applied drives due to Floquet dynamics.
Such phenomena can have substantial impact and introduce additional dynamics in those schemes utilizing continuous monitoring of the mechanical oscillator with multiple tones, such as backaction-evading measurement~\cite{suh_mechanically_2014,shomroni2018}, mechanical quantum squeezing~\cite{kronwald_arbitrarily_2013,wollman_quantum_2015,lecocq_quantum_2015,pirkkalainen_squeezing_2015,lei2016}
and dissipative optical squeezing~\cite{kronwald_dissipative_2014}, entanglement of two mechanical oscillators~\cite{woolley_two-mode_2014,ockeloen-korppi_stabilized_2018}, optomechanical non-reciprocity~\cite{bernier_nonreciprocal_2017,peterson_demonstration_2017,Xu_2018}.
The observed Floquet dynamics can be exploited for future optomechanical Floquet engineering with time periodic modulation, such as generation of optomechanical topological phases~\cite{schmidt2015,mathew2018} and continuous variable quantum information~\cite{schmidt2012}.

\section*{Data availability}
The code and data used to produce the plots within this paper will be
made available via zenodo.org upon publication.

\begin{acknowledgments}
  The authors acknowledge discussions with C.~Galland, V.~Sudhir and~H.~R.~Guo.
  LQ acknowledges support by Swiss National Science Foundation under grant No.~163387.
  IS acknowledges support by the European Union's Horizon 2020 research and innovation programme under Marie Sko\l{}odowka-Curie IF grant agreement No.~709147 (GeNoSOS).
  DM acknowledges support by the UK Engineering and Physical Sciences Research Council (EPSRC) under Grant No.~EP/M506485/1.
  AN acknowledges a University Research Fellowship from the Royal Society and support from the Winton Programme for the Physics of Sustainability.
  TJK acknowledges financial support from an ERC AdG (QuREM).
  This work was supported by the SNF, the NCCR Quantum Science and Technology (QSIT), and the European Union's Horizon 2020 research and innovation programme under grant agreement No.~732894 (FET Proactive HOT).
  All samples were fabricated in the Center of MicroNanoTechnology (CMi) at EPFL.
\end{acknowledgments}

\appendix

\section{The heterodyne spectrum}\label{app:heterodyne_spectrum}
In this Appendix we give more details on heterodyne detection and the resulting spectrum.
During balanced heterodyne detection, we mix the signal with a strong local oscillator of amplitude $\alpha_{\text{LO}}\exp(-i\omega_{\text{LO}}t)$, and thus detect the heterodyne photocurrent \cite{BowenMilburn2015}
\begin{equation}
  \hat I_{\text{out}}(t)=e^{i\Delta_{\text{LO}}t}\alpha^*_{\text{LO}}\delta\hat a_{\text{out}}(t)+e^{-i\Delta_{\text{LO}}t}\alpha_{\text{LO}}\delta\hat a_{\text{out}}\dagg(t),
  \label{eq:heterodyne_current}
\end{equation}
where we have assumed that the local oscillator is detuned from the cavity by $\Delta\s{LO}=\omega\s{LO}-\omegacav$, and the output operators are in a frame rotating with the cavity.
The (time-averaged) autocorrelator of the signal is the heterodyne spectrum as quoted in the text and repeated here for convenience
\begin{equation}
  	\bar S_I(\omega)=\frac12\int_{-\infty}^\infty\langle\{\overline{\hat I_{\text{out}}(t+t'),\hat I_{\text{out}}(t')}\}\rangle e^{i\omega t}dt,
  \label{eq:heterodyne_spectrum}
\end{equation}
where the overbar denotes averaging with respect to $t'$.
Invoking commutation relations we normal-order the operators in Eq.~\eqref{eq:heterodyne_spectrum}
\begin{equation}
  \begin{aligned}
	&\left\{ \overline{\hat I_{\text{out}}(t+t'),\hat I_{\text{out}}(t')} \right\}=|\alpha_{\text{LO}}|^2\left[\delta(t)+\right. \\
	  &\left.\overline{\delta\hat a_{\text{out}}\dagg(t+t')\delta\hat a_{\text{out}}(t')}e^{-i\Delta_{\text{LO}}t}
	+\overline{\delta\hat a_{\text{out}}\dagg(t')\delta\hat a_{\text{out}}(t+t')}e^{i\Delta_{\text{LO}}t}\right].
  \end{aligned}
  \label{eq:normal-ordering}
\end{equation}
The first term is vacuum noise, whereas the second and third term give the normalized output spectrum around $\pm\Delta_{\text{LO}}$.
Note that the anomalous terms have disappeared from the right-hand side of Eq.~\eqref{eq:normal-ordering} as a result of the averaging over $t'$.
Keeping only the positive frequency term and normalizing to the shot noise floor, we have $\bar S_I(\omega+\Delta_{\text{LO}})=1+\eta S^N(\omega)$, where $\eta$ is the overall detection efficiency \cite{gardiner2004quantum,BowenMilburn2015}, and the normal-ordered spectrum
\begin{equation}
  S^N(\omega)=\int_{-\infty}^\infty\overline{\langle \delta\hat a_{\text{out}}\dagg(t+t')\delta\hat a_{\text{out}}(t')\rangle}e^{i\omega t}dt.
  \label{eq:normal-ordered_spectrum}
\end{equation}

In order to determine $\delta\hat a_{\text{out}}$ we use the quantum Langevin equations \eqref{eq:langevin_eoms}
together with the input-output relations $\delta\hat a_{\text{out}}=\delta\hat a_{\text{in}}-\sqrt{\kappa_{\text{ex}}}\delta\hat a$.
For just one tone on the red sideband, we have (in the rotating-wave approximation)
$\delta\hat a=\chi\s{opt}(\omega)(ig\delta\hat b+\sqrt{\kappa}\delta\hat a_{\text{in}})$ [susceptibilities defined in Eq.~\eqref{eq:susceptibilities}],
such that
\begin{equation}
  \bar S_I(\omega+\Delta_{\text{LO}})=1+\eta\Gamma_m^2\kappa^2\C_{\text{red}}|\chi\s{opt}(\omega)|^2|\chi_m(\omega)|^2\bar n.
  \label{eq:red_sideband_spectrum}
\end{equation}
Conversely, a tone on the blue sideband results in $\delta\hat a=\chi\s{opt}(\omega)ig\delta\hat b\dagg+\sqrt{\kappa}\delta\hat a_{\text{in}}$ and thus a sideband of weight $\bar n+1$
\begin{equation}
  \bar S_I(\omega+\Delta_{\text{LO}})=1+\eta\Gamma_m^2\kappa^2\C_{\text{blue}}|\chi\s{opt}(\omega)|^2|\chi_m(\omega)|^2(\bar n+1).
  \label{eq:blue_sideband_spectrum}
\end{equation}
In each case, $\tilde\Delta$ is the residual detuning of the tone from the respective sideband.
In this calculation the sideband asymmetry ($\propto\bar n$ vs.\ $\propto\bar n+1$) arises due to the fact that for red-detuned probing the normal-ordered optical output correlator is proportional to $\langle\hat b\s{in}\dagg\hat b\s{in}\rangle$, whereas for blue-detuned probing it is proportional to $\langle\hat b\s{in}\hat b\s{in}\dagg\rangle$.

If both tones are present and weak enough such that their backaction can be neglected, the spectrum will be the sum of both contributions (except the shot noise level, which stays unchanged), as long as the sidebands remain well-separated from each other, which can be achieved by detuning the tones from each other by much more than the effective mechanical linewidth $\Gamma\s{tot}$.
Further including classical laser noise, but neglecting the frequency dependence of the optical susceptibility yields Eq.~(A23) in Ref.~\cite{sudhir_appearance_2017}.
An additional strong cooling tone will contribute optical damping to the mechanical oscillator, and generate another sideband, which yields Eq.~\eqref{eq:ideal_theory}.

In order to calculate the spectrum in the presence of the Kerr effect, we have to otain the spectrum of a modulated system. Recall that in our experiment, both the optomechanical coupling and the cavity frequency oscillate as a consequence of the oscillating intracavity field.
This leads to Langevin equations with an explicit time-dependence~\eqref{eq:langevin_eoms} of the general form (where $\vec x$ is a vector of operators)
\begin{equation}
  \dot{\vec x}(t)=\matr A(t)\vec x(t)+\matr L\vec x_{\text{in}}(t).
  \label{eq:periodic_langevin}
\end{equation}
As done in the main text, this can be solved perturbatively by introducing Fourier modes $\vec x(t)=\exp(i\Omega_{\text{mod}}nt)\vec x^{(n)}(t)$, where $\Omega_{\text{mod}}$ is the fundamental frequency of the modulation in $\matr A(t)$.
One can then show that the normal-ordered time-averaged spectrum~\eqref{eq:normal-ordered_spectrum} is given by Eq.~\eqref{eq:measured_spectrum}.

\section{Sideband Cooling, Absorption Heating, and Effect of the Buffer Gas}
\label{sec:sideband_cooling}

An important characterization of our system is done by measuring the performance of sideband cooling~\cite{aspelmeyer_cavity_2014}.
A single tone is applied at the lower mechanical sideband,
$\Delta=-\Omega_m$.
In the well-resolved sideband regime, dynamical backaction provides additional damping $\Gamma\s{opt} =\C\Gamma_m$
where $\C=\C_0n_c$ is the optomechanical cooperativity for $n_c$ mean intracavity photons and single-photon cooperativity $\C_0=4g_0^2/(\kappa\Gamma_m)$.
The mean thermal occupation of the oscillator is lowered to
\begin{equation}
  \bar{n} = \frac{\Gamma_m \bar n\s{th} + \Gamma\s{opt}n\s{qbl}}{\Gamma_m+\Gamma\s{opt}}
  = \frac{\bar n\s{th}+\C n\s{qbl}}{1+\C},
\end{equation}
where $\bar n\s{th}$ is the mean thermal occupation of the environment
and $n\s{qbl} = \kappa^2/(16\Omega_m^2)\sim 6\times 10^{-3}$ is the quantum limit of sideband cooling~\cite{wilson-rae_theory_2007,marquardt_quantum_2007}, and will be neglected here.

The simplest treatment of optical absorption heating is provided by the linear model $\bar n\s{th} \rightarrow \bar n\s{th} + \heating n_c$, where $\heating$ is the extraneous heating of the bath in terms of oscillator phonons per intracavity photon.
Within this model, we have
\begin{equation}
  \bar n = \frac{\bar n\s{th}+\heating n_c}{1+\C},
  \label{eq:heating_model}
\end{equation}
the asymptotic quantum limit $n\s{qbl}$ is effectively replaced by the heating limit of $\heating/\C_0$ phonons.
In this model, we find the signal-noise-ratio (SNR), defined as the peak of $\bar S_I(\omega)$ relative to the noise floor
\begin{equation}
  \mathrm{SNR} = \frac{4\eta\C\bar n}{1+\C}=4\eta\frac{(\bar n\s{th}+[\heating/\C_0]\C)\C}{(1+\C)^2}.
  \label{eq:snr_model}
\end{equation}
The first equality can be obtained from Eq.~\eqref{eq:red_sideband_spectrum}, the second from Eq.~\eqref{eq:heating_model}.

\begin{figure*}[thb]
  \includegraphics[scale=1]{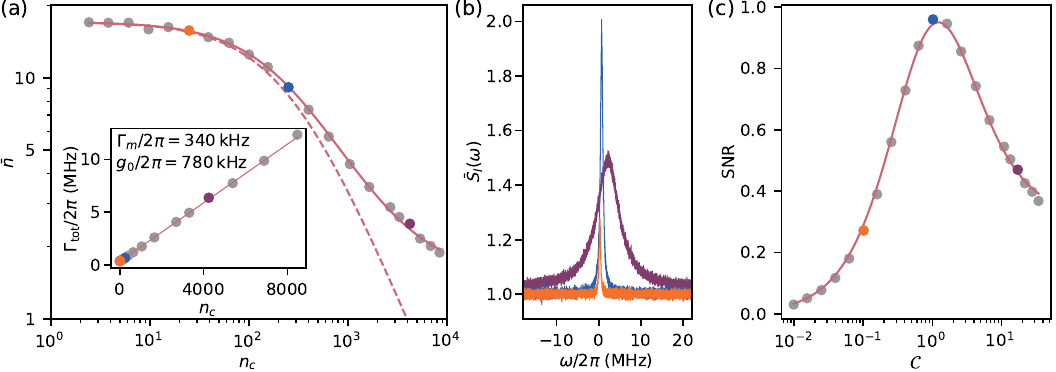}
  \caption{\textbf{Sideband cooling of the nanomechanical oscillator with a single cooling tone.}
  In this instance, with $4.5\unit{K}$ cryostat temperature and
  $160\unit{mbar}$ buffer gas pressure, the oscillator was cooled to occupancy $\bar n=1.9$, i.e.~34\% ground-state occupation.
  (a)~Mechanical occupation $\bar n$ derived from the area of the noise spectra vs.~intracavity photon number $n_c$.
  $\bar n$ is anchored to the cryostat temperature, $\bar n\s{th} = 17$, at lowest cooling power.
  The solid line is a fit to the simple heating model~\eqref{eq:heating_model} with $\heating$ and $\bar n\s{th}$ as free parameters, and the dashed line shows cooling in absence of absorption heating.
  The inset shows total mechanical damping $\Gamma\s{tot}$ vs.~$n_c$, with a linear fit used to extract
  the bare mechanical damping rate $\Gamma_m$ (intrinsic damping $\Gamma\s{int}$ and the gas damping) in absence of dynamical backaction
  and the optomechanical coupling rate $g_0$.
  (b)~Noise spectra $\bar S_I(\omega)$, normalized to the shot noise level, at three different points highlighted with respective colors in the main panel.
  (c)~Signal-to-noise ratio (SNR) vs.~the cooperativity $\C$ with a fit to the model~\eqref{eq:snr_model} with overall efficiency $\eta$ and heating~$\heating$ as free parameters.
  The highlighted points are the same as in~(b).
}
\label{fig:sideband_cooling}
\end{figure*}

Figure~\ref{fig:sideband_cooling}a shows an example sideband cooling at cryostat temperature of~$4.5\unit{K}$ with the buffer gas pressure at $160\unit{mbar}$.
In order to calibrate $\bar n$, we assume the heating is negligible at the point of lowest power (with $n_c\sim 2$ intracavity photons) and the oscillator is completely thermalized to the cryostat, $\bar n=\bar n\s{th}$.
We infer $\bar n$ from the area of the noise spectra (Fig.~\ref{fig:sideband_cooling}b).
The SNR of the spectra is shown in Fig.~\ref{fig:sideband_cooling}c.
Fitting the data of Fig.~\ref{fig:sideband_cooling}a,c using Eqs.~\eqref{eq:heating_model} and~\eqref{eq:snr_model} respectively, we obtain extraneous heating of $\heating=6.3\times 10^{-3}=1.5\,\C_0$ and overall efficiency $\eta=0.04$ in agreement with Sec.~\ref{sec:ExprResults}.

\begin{figure}[thb]
  \includegraphics[scale=1]{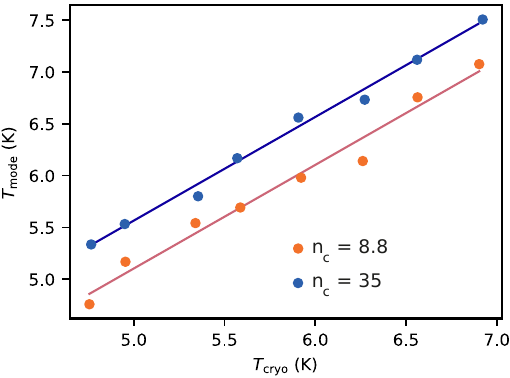}
  \caption{\textbf{Laser power induced heating of the mechanical mode.}
  Mechanical mode temperature versus cryostat temperature at two different pumping powers which correspond to intracavity photon number $n_c=8.8$ and $n_c=35$, where the sample pressure is kept at $25\unit{mbar}$}
  \label{fig:T_Mode_T_Cryo}
\end{figure}

Working in a buffer gas environment enables us to substantially improve the thermalization of our optomechanical system.
In Fig.~\ref{fig:T_Mode_T_Cryo} we show the mechanical mode temperature as a function of the temperature of the cryostat for two different pump powers,  corresponding to mean intracavity photon number of 8.8 and 35, keeping the buffer gas pressure constant at $25\unit{mbar}$.
The mechanical mode temperature is consistently higher for the higher input power, giving clear indication of optical absorption heating.

\begin{figure}[thb]
  \includegraphics[scale=1]{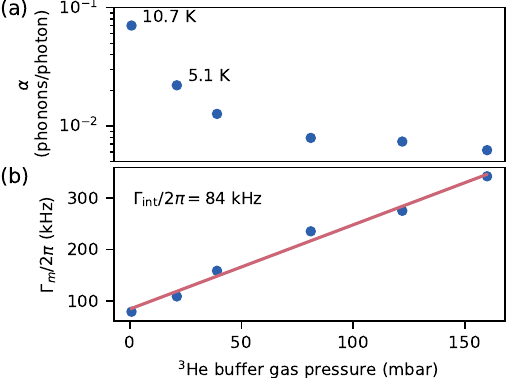}
  \caption{\textbf{Effect of the buffer $^3$He gas on sideband cooling performance.}
  (a)~Mechanical damping showing increase with gas pressure.
  The linear field yields intrinsic damping $\Gamma\s{int}/2\pi=84\unit{kHz}$.
  (b)~Oscillator heating $\heating$
  showing decrease with gas pressure.
  Measurements were done at cryostat temperature of $4.5\unit{K}$ except for
  low pressures, where insufficient thermalization leads to elevated bath temperatures, as indicated in~(a).
  At pressure of $\sim 0\unit{mbar}$, absorption heating leads to minimum $\bar{n}$ of 5.7, despite the low $\Gamma_m$.
}
\label{fig:power_sweep_pressure}
\end{figure}

To more accurately quantify the optical absorption heating rate, we studied this effect by performing sideband cooling measurements at different buffer gas pressures, with the results shown in Fig.~\ref{fig:power_sweep_pressure}.
As expected, increasing the pressure leads to a decrease of the extraneous heating $\heating$ of the mechanical mode (Fig.~\ref{fig:power_sweep_pressure}a), while at the same time introducing additional mechanical damping due to the gas (Fig.~\ref{fig:power_sweep_pressure}b).
The combined effect of gas damping and thermalization results in theoretical cooling limits $\heating/\C_0$ of 1.3--1.5 phonons
in excellent agreement with Sec.~\ref{sec:ExprResults}, using more accurate sideband asymmetry calibration.

\section{Design and Fabrication}
\label{sec:fabrication}

\begin{figure}[thb]
  \includegraphics[scale=1]{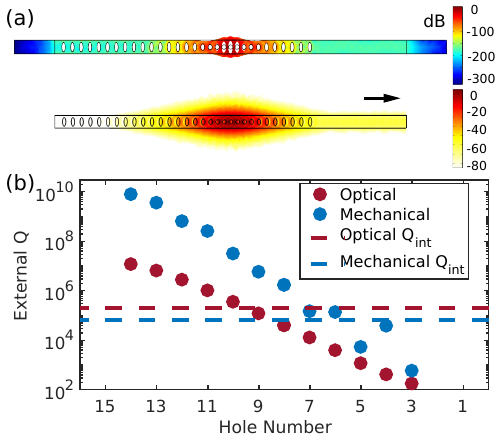}
  \caption{\textbf{Design of the front input coupling mirror of the optomechanical crystal (OMC).}
  (a)~Simulated mechanical (top) and optical (bottom) field distribution of the OMC.
  (b)~Simulated optical and mechanical external $Q$ vs.~number of holes of the front mirror, adjacent the defect section of OMC.
  When the number of holes is increased, both optical (red dot) and mechanical (blue dot) external $Q$ increases.
  The dashed lines correspond to the measured intrinsic $Q$ of the optical (red) and mechanical (blue) mode.
}
\label{fig:design}
\end{figure}

Our optomechanical system is an optomechanical crystal (OMC) \cite{eichenfield_optomechanical_2009} similar in design to that reported previously \cite{chan_optimized_2012} (Fig.~\ref{fig:experimental}a,b).
A periodic pattern of holes in a silicon nanobeam creates a bandgap preventing propagation of both photons and acoustic phonons.
A defect in the pattern is designed to co-localize photons and phonons which are coupled by radiation pressure and electrostrictive forces (Fig.~\ref{fig:experimental}a, inset).
By controlling the number of holes on each side of the defect, a single-sided geometry is realized,
where the front input coupling mirror has less holes than the high-reflectivity mirror.
Figure~\ref{fig:design} shows the design for the OMC. In Fig.~\ref{fig:design}a, the simulated mechanical and optical mode field distribution are presented.
In such a single-sided cavity, the optical field can be efficiently coupled out while the mechanical mode is maintained localized.
Figure~\ref{fig:design}b shows the simulated external $Q$ of the optical and mechanical modes vs.~the hole number of the front mirror.
Increasing the number of holes in the front mirror leads to an increased external $Q$ of both optical and mechanical mode.
However in practice, the mechanical $Q$ is much lower, limited by material and fabrication imperfections, and hence not affected by hole number.
By comparing to the measured intrinsic optical $Q$, a proper hole number for the front mirror is chosen in the actual design, while the hole number of the high-reflectivity mirror is kept fixed.

Beyond the patterned section, the nanobeam is extended to introduce a coupling waveguide, into which light can be evanescently coupled by positioning a tapered optical fiber of diameter $\sim 1\unit{\mu m}$.
A bend in the nanobeam offsets the coupling waveguide such that the tapered fiber does not touch the optomechanical cavity.
In order to reduce the scattering loss, the $529\unit{nm}$ width of the nanobeam is tapered down to $280\unit{nm}$ adiabatically.
To maximize the coupling, a $3.05\unit{\mu m}$ long nanobeam mode convertor with a width of $280\unit{nm}$ is added to convert the light from the tapered fiber.
Due to phase mismatch between the waveguide and the tapered fiber, we obtain typical single-pass waveguide coupling efficiencies $\eta\s{wg}>50\%$ experimentally,
while the FEM simulation gives  $\eta\s{wg} > 90\%$.

The optomechanical crystal is fabricated on a silicon on insulator wafer (Soitec) with a device layer of $220\unit{nm}$ and $2\unit{\mu m}$ buried oxide layer.
We pattern our chips with an electron beam lithography using ZEP as resist.
The structures are transferred to the device layer following a reactive ion etching (RIE) using a SF$_6$/C$_4$F$_8$ plasma.
To open an area for taper fiber coupling, an additional photolithography and RIE are applied to create a mesa structure.
After removing all the resist, the device layer is undercut in diluted 10\% hydrofluoric acid.
Following an additional Piranha (a mixture of sulfuric acid and hydrogen peroxide) cleaning step to remove organic residuals,
the sample is finally dipped into 2\% hydrofluoric acid to terminate the device surface with hydrogen atoms.
The sample is then immediately mounted on the probe for cryogenic characterization.

\section{Experimental System Details}
\label{sec:expr_details}
Cryogenically cooled optomechanical crystals are promising platforms for various quantum optomechanical
experiments~\cite{riedinger_non-classical_2016,hong_hanbury_2017,riedinger_remote_2018}, mainly due to their GHz-scale mechanical frequencies, which for temperatures in the
millikelvin range achievable by dilution refrigerators, puts them very close to the ground state
$\bar{n}\sim k_B T/\hbar\Omega_m \ll 1$. However, the vacuum environment combined with the nanobeam
geometry and falling thermal conductivity of silicon at cryogenic temperatures, lead to large heating due to
optical absorption~\cite{meenehan_silicon_2014}.
This severely limits both the minimum achievable phonon occupation
and the optomechanical cooperativity, set by the number of intracavity photons.

We choose instead to use a $^3$He buffer gas cryostat (Oxford Instruments HelioxTL).
The cold buffer gas provides a thermalization channel thereby reducing the heating effect, and was employed in prior work on quantum coherent coupling~\cite{verhagen_quantum-coherent_2012}.
The pressure of the buffer gas is controlled through a sorption pump, and enables us to introduce lower heating at the expense of higher mechanical damping due to the gas.
A study of this trade-off is presented in Appendix~\ref{sec:sideband_cooling}.

The experimental setup is shown in Fig.~\ref{fig:experimental}d.
For the sideband asymmetry setup Fig.~\ref{fig:experimental}c, the red-detuned probe, local oscillator and blue-detuned probe, separated in frequency by $\sim\Omega_m$, are generated by three lasers phase-locked in a chain.
The cooling tone is derived from the red-detuned probe laser by an
acousto-optical frequency shifter (AOM).
For the dual cooling tone setup of Fig.~\ref{fig:dual_tone_all}a, the cooling tone is eliminated, and the former blue-detuned probe is tuned close to the red-detuned probe, with continuous control of the dual tone separation. Note that we have observed the same results if deriving both tones from the same laser, via AOM.
The three beams are attenuated to the desired power level and combined in free-space with the same polarization into a single-mode fiber.
Once in the fiber, we may monitor each beam by blocking the two others.
The power stability throughout the experiment is $\sim 1\%$.
A fiber-optic circulator feeds the reflected light to the detection stage.
The reflected signal is mixed with a strong ($\sim 7\unit{mW}$) local oscillator in a balanced heterodyne detection scheme, and the noise power spectral density of the subtracted photocurrent is analyzed using a spectrum analyzer.

To achieve the desired tuning of the tones relative to the optical resonance, we temporarily switch the reflected light to a coherent response setup, employing a network analyzer driving a phase modulator on the master laser~\cite{weis_optomechanically_2010}.
The acquired response is fit with a theoretical curve, yielding the actual detuning $\Delta$ as well as cavity linewidth $\kappa$.
We estimate our tuning accuracy to be $\pm 2\pi\times 20\unit{MHz}$.
All asymmetry measurements presented in this paper are corrected based on measured detuning.

\section{Excess Laser Noise}
\label{sec:excess_noise}
External cavity diode lasers are well-known to have an excess noise in the GHz range, due to the damped relaxation oscillation caused by carrier population dynamics.
This is the main contribution for GHz excess noise for diode lasers. As this frequency is close to the mechanical frequency, it is necessary to quantify the phase noise of the diode laser that we are using \cite{safavi-naeini_laser_2013, kippenberg_phase_2013}.
The presence of the excess laser phase noise leads to a limited sideband cooling performance.

On the other hand, the effect of excess laser noise on the sideband asymmetry has been extensively studied \cite{jayich_cryogenic_2012,safavi-naeini_laser_2013,sudhir_appearance_2017} in both sideband resolved and sideband unresolved regimes.
In a heterodyne detection scheme, the sideband imbalance is due to the correlation between imprecision and backaction from the probing tones.
This is true also when the probing tones are not quantum limited, for example a laser with phase noise or a microwave source with thermal occupation.
This can therefore lead to an artificially increased asymmetry. As a consequence it is imperative that the lasers are quantum noise limited in amplitude and phase quadrature.

Here we give an estimation of the effect from the measured excess laser noise on sideband asymmetry.
We take into consideration of the excess noise in the amplitude and phase quadratures, defined as $C_{qq}$ and $C_{pp}$ respectively, as in \cite{sudhir_appearance_2017}.

To quantify the excess noise in the amplitude quadrature $C_{qq}$,
we measure the power spectral density $\bar{S}\s{p}$ of the photocurrent at mechanical frequency $\Omega_m$ to the incident optical power through direct detection, where
$\bar{S}_p=2(\hbar\omega_L)^2\avg{\dot n}(1+2C_{qq})=\bar{S}_p^{\mathrm{shot}}+\bar{S}_p^{\mathrm{ex}}$.
The excess amplitude noise leads to a deviation of
$\bar{S}_p^{\mathrm{ex}}=4(\hbar\omega_L)^2\avg{\dot n}C_{qq}$
from the shot noise power spectral density,
which becomes visible at high optical power (100s $\mathrm{\mu W}$ to few mW).
At the mechanical frequency $\Omega_m/2\pi=5.3\unit{GHz}$, for typical probing power (0.5--$2\unit{\mu W}$) we use in the sideband asymmetry measurement,
$C_{qq}(\Omega_m)<10^{-4}$, which is negligible.

\begin{figure*}[tbh]
  \includegraphics[scale=1]{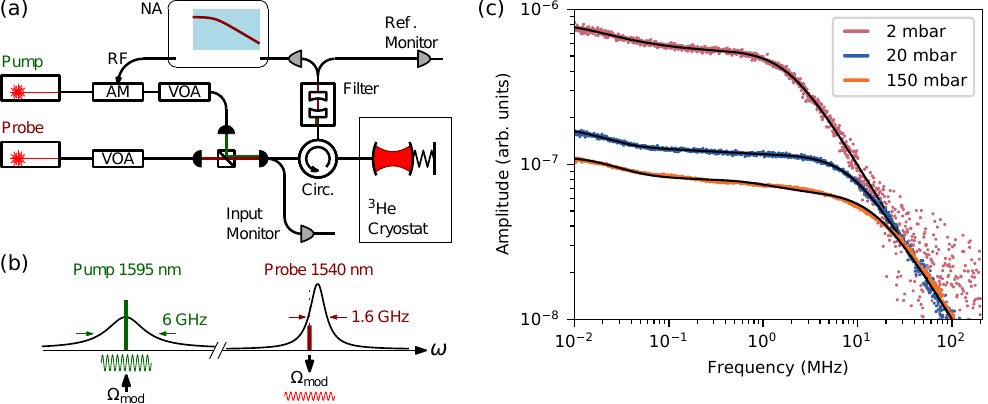}
  \caption{\textbf{Cavity frequency response measurement.}
  (a)~Experimental setup.
  AM, amplitude modulator.
  (b)~Measurement scheme.
  (c)~Response measurement at different pressures fitted with a three low-pass model.
}
\label{fig:response}
\end{figure*}

The excess noise in the phase quadrature $C_{pp}$ can be modeled as
$C_{pp}=\avg{\dot n}S^{\mathrm{ex}}_{\phi\phi}$, where $S^{\mathrm{ex}}_{\phi\phi}$ is the phase noise spectral density of the laser.
The phase noise of the laser is characterized with a narrow filter cavity, which transduces the phase fluctuation of the input field to amplitude fluctuation.
For the laser (Toptica CTL) at typical current $300\unit{mA}$, the relaxation oscillation frequency is around $1.94\unit{GHz}$ with a frequency noise spectral density of $S_{\omega\omega}(\Omega\s{relax})=2\times10^6\unit{rad^2 Hz}$.
At the mechanical frequency, the characterized frequency noise spectral density
$S_{\omega\omega}(\Omega_m)$ is less than $10^5\unit{rad^2 Hz}$.
For the typical probing power, $C_{pp}$ is below $10^{-3}$.

As a result, the excess laser noise in both quadratures at typical probing power at the Fourier frequency of $5.3\unit{GHz}$ are negligible compared to the quantum fluctuations of the light.

\section{Frequency Response Measurement}
\label{sec:frequency_response}

To verify our theory further, and to discern contributions from different Kerr-type nonlinearities, we performed a cavity frequency response measurement~\cite{rokhsari_observation_2005,schliesser_radiation_2006}, shown in Fig.~\ref{fig:response}.
In this pump-probe scheme, a pump laser is applied to a secondary wide optical resonance at $1595\unit{nm}$,
while a probe laser is tuned to the slope of the main optical resonance at $1540\unit{nm}$ (Fig.~\ref{fig:response}b)
The pump laser is amplitude-modulated with variable frequency $\Omega$ and the corresponding shift of the cavity resonance is monitored via the probe laser.
A bandpass optical filter at the output removes the reflected pump light to avoid cross-talk from the pump modulation.

The frequency response curve corresponds to a combination of low-pass filter behavior, attributed to different physical mechanisms that dominate at different frequencies~\cite{schliesser_cavity_2009}. In our case we found it best described by three low-pass components
\begin{equation}
  \delta\omega= \left(\frac{a_1}{1+i\Omega/\Omega_1}+\frac{a_2}{1+i\Omega/\Omega_2}+\frac{a_3}{1+i\Omega/\Omega_3}\right)\delta n_c
  \label{eq:three_filter}
\end{equation}
with $\delta\omega$ the cavity frequency modulation, $\delta n_c$ the intracavity photon number modulation, and $\Omega_i$ and $a_i$ the bandwidths and amplitudes, respectively, of the three filters.
Note that neither the intrinsic Kerr effect plateau, nor the mechanical oscillator response, were observed in our measurements, which were dominated by detector noise at GHz frequencies.

Figure~\ref{fig:response}c shows our cavity frequency response for different buffer gas pressures, with the extracted bandwidths summarized in Table~\ref{tab:response}, from which several conclusions can be drawn.
First, at $150\unit{mbar}$ the highest frequency $\Omega_3$ around $15\unit{MHz}$, agrees with the bandwidth of the Kerr-type effect studied in Sec.~\ref{sec:ExprResults}, which reinforces our observations (the lower frequencies are too low to be observed).
Second, all bandwidths increase with pressure, indicating the role of the buffer gas in improving the thermalization of the cavity. In the range 2--$150\unit{mbar}$, The higher bandwidths $\Omega_{2,3}$ increase by an order of magnitude.
Thus, the cavity response measurement plays a key role in understanding the Kerr-type nonlinearities studied in this work.
\setlength{\tabcolsep}{10pt}
\renewcommand{\arraystretch}{1.25}
\begin{table}[h]
  \begin{tabular}{llll}
    Pressure (\unit{mbar}) 		   & 2 & 20 & 150 \\ \hline
    $\Omega_{1}/2\pi (\unit{kHz})$ &	16.5 & 17.6 & 24.6\\ 
    $\Omega_{2}/2\pi (\unit{MHz})$ & 	0.07 & 0.30 & 1.10\\ 
    $\Omega_{3}/2\pi (\unit{MHz})$ & 	1.84 & 8.62 & 15.4\\ 
  \end{tabular}
  \caption{Bandwidth from fitting the response measurements of Fig.~\ref{fig:response} with the three low-pass model~\eqref{eq:three_filter}.}
  \label{tab:response}
\end{table}

%

\end{document}